\documentclass[journal]{IEEEtran}

\ifCLASSINFOpdf
\else
  \usepackage[dvips]{graphicx}
\fi

\usepackage{url}
\usepackage{color, soul}
\usepackage{xcolor}
\usepackage{cite}
\usepackage[colorlinks,linkcolor=black]{hyperref}
\usepackage{multirow}
\usepackage{amsmath}
\usepackage{bm}
\usepackage{graphicx}
\usepackage[keeplastbox]{flushend}
\usepackage{comment}
\usepackage{longtable}
\usepackage{xparse}
\usepackage{booktabs}
\usepackage{subfigure}
\usepackage{caption}
\usepackage{array}
\usepackage{ulem}

\usepackage{lettrine}

\newcommand{\tabincell}[2]{\begin{tabular}{@{}#1@{}}#2\end{tabular}}
\usepackage{makecell}

\begin{document}

\title{Expressive TTS Training with Frame and Style Reconstruction Loss}

\author{Rui Liu, \IEEEmembership{Member, IEEE},  Berrak Sisman, \IEEEmembership{Member, IEEE}, Guanglai Gao,\\ Haizhou Li, \IEEEmembership{Fellow, IEEE}

\vspace{-2mm}
\thanks{This paper is submitted on 19 July 2020 for review. This research is funded by SUTD Start-up Grant Artificial Intelligence for Human Voice Conversion (SRG ISTD 2020 158) and SUTD AI Grant, titled 'The Understanding and Synthesis of Expressive Speech by AI'. The research by Haizhou Li is supported by the National Research Foundation, Singapore under its AI Singapore Programme (Award No: AISG-GC-2019-002) and (Award No: AISG-100E-2018-006), and its National Robotics Programme (Grant No. 192 25 00054), and by RIE2020 Advanced Manufacturing and Engineering Programmatic Grants A1687b0033, and A18A2b0046.}



\thanks{Rui Liu is with Singapore University of Technology and Design (SUTD) and National University of Singapore (e-mail: liurui\_imu@163.com). }
\thanks{Berrak Sisman is with Singapore University of Technology and Design (SUTD), Singapore (e-mail: berraksisman@u.nus.edu). }
\thanks{Guanglai Gao is with the Department of Computer Science, Inner Mongolia University, China.}
\thanks{Haizhou Li is with the Department of Electrical and Computer Engineering, National University of Singapore. He is also with University of Bremen, Faculty 3 Computer Science / Mathematics, Enrique-Schmidt-Str. 5
Cartesium, 28359 Bremen, Germany (e-mail: haizhou.li@nus.edu.sg). }}

\markboth{PREPRINT MANUSCRIPT OF IEEE/ACM TRANSACTIONS ON AUDIO, SPEECH, AND LANGUAGE PROCESSING}%
{Shell \MakeLowercase{\textit{et al.}}: Bare Demo of IEEEtran.cls for IEEE Journals}

\maketitle

\begin{abstract}
We propose a novel training strategy for Tacotron-based text-to-speech (TTS) system that improves the speech styling at utterance level.  One of the key challenges in prosody modeling is the lack of reference that makes explicit modeling difficult. The proposed technique doesn't require prosody annotations from training data. It doesn't attempt to model prosody explicitly either, but rather encodes the association between input text and its prosody styles using a Tacotron-based TTS framework. 
This study marks a departure from the style token paradigm where prosody is explicitly modeled by a bank of prosody embeddings.
It adopts a combination of two objective functions: 1) frame level reconstruction loss, that is calculated between the synthesized and target spectral features; 2) utterance level style reconstruction loss, that is calculated between the deep style features of synthesized and target speech. The style reconstruction loss is formulated as a perceptual loss to ensure that utterance level speech style is taken into consideration during training. 
Experiments show that the proposed training strategy achieves remarkable performance and outperforms the state-of-the-art baseline in both naturalness and expressiveness. To our best knowledge, this is the first study to incorporate utterance level perceptual quality as a loss function into Tacotron training for improved expressiveness.
 
\end{abstract}


\IEEEpeerreviewmaketitle

\vspace{-3mm}
\section{Introduction}

\IEEEPARstart{W}{ith} the advent of deep learning, neural TTS has shown many advantages over the conventional TTS techniques \cite{tokuda2013speech, ze2013statistical,liu2017mongolian}. For example, encoder-decoder architecture with attention mechanism, such as  Tacotron~\cite{wang2017tacotron,shen2018natural,liu2020wavetts,lee2019robust}, has consistently achieved high voice quality. The key idea is to integrate the conventional TTS pipeline \cite{hunt1996unit,tokuda2002hmm} into an unified framework that learns  sequence-to-sequence mapping from text to a sequence of acoustic features~\cite{lee2019robust,chung2019semi,He2019,Luong2019,liu2019teacher}. Furthermore, together with a neural vocoder \cite{hayashi2017investigation,shen2018natural,chen2018high,Okamoto2019, berrak_is18, berrak-journal, sisman2018adaptive}, neural TTS  generates natural-sounding and human-like speech which achieves state-of-the-art performance.  Despite the progress, the expressiveness of the synthesized speech remains to be improved. 

Speech conveys information not only through phonetic content, but also through its prosody.  Speech prosody can affect syntactic and semantic interpretation of an utterance \cite{hirschberg2004pragmatics}, that is called linguistic prosody. Speech prosody is also used to display one's emotional state, that is referred to as affective prosody.
Both linguistic prosody and affective prosody are manifested over a segment of speech beyond short-time speech frame. Linguistically, speech prosody in general refers to stress, intonation, and rhythm in spoken words, phrases, and sentences. As speech prosody is the result of the interplay of multiple speech properties, it is not easy to define speech prosody by a simple labeling scheme~\cite{Luong,lin2019investigation,hodariusing,zhao2020improved,hodari2020perception}. Even if a labeling scheme is possible~\cite{Silverman1992TOBIAS,taylor1998assigning}, a set of discrete labels may not be sufficient to describe the entire continuum of speech prosody. 

Besides naturalness, one of the factors that differentiate human speech from today's synthesized speech is their expressiveness.
Prosody is one of the defining features of expressiveness that makes speech lively. Several recent studies successfully improve the expressiveness of Tacotron TTS framework\cite{ wang2018style,Stanton2018Predicting,skerry2018towards,sun2020fully,sun2020generating}. The idea is to learn latent prosody embedding, i.e. style token, from training data~\cite{wang2018style}. At run-time, the style token can be used to predict the speech style from text~\cite{Stanton2018Predicting}, or to transfer the speech style from a reference utterance to target~\cite{skerry2018towards}. 
It is observed that such speech styling is effective and  consistently improves speech quality. Sun et al. \cite{sun2020fully, sun2020generating} further study a hierarchical, fine-grained and interpretable latent variable model for prosody rendering. The studies show that precise control of the prosody style leads to improvement of prosody expressiveness in the Tacotron TTS framework. However, several issues have hindered the effectiveness of above prosody modeling techniques. 

First, the latent embedding space of prosody is learnt in an unsupervised manner, where the style is defined as anything but speaker identity and phonetic content in speech. We note that many different styles co-exist in speech. Some are speaker dependent, such as accent and idiolect, others are speaker independent such as prosodic phrasing, lexical stress and prosodic stress. There is no guarantee that such latent embedding space of style represents only the intended prosody. Second, while the techniques don't require the prosody annotations on training data, they require a reference speech or a manual selection of style token~\cite{wang2018style} in order to explicitly control the style of output speech during run-time inference. While it is possible to automate the style token selection \cite{Stanton2018Predicting}, a correct prediction of style token is subject to both the design of the style token dictionary, and the run-time style token prediction algorithm. Third, the style token dictionary in Tacotron is trained from a collection of speech utterances to represent a large range of acoustic expressiveness for a speaker or an audiobook\cite{wang2018style}. It is not intended to provide differential prosodic details at phrase or utterance level. It is desirable for Tacotron system to learn to automate the prosody styling in response to input text at run-time, that will be the focus of this paper.

To address the above issues, we believe that Tacotron training should minimize frame level reconstruction loss~\cite{wang2017tacotron,shen2018natural} and utterance level perceptual loss at the same time. Perceptual loss is first proposed for image stylization and synthesis \cite{ dosovitskiy2016generating,johnson2016perceptual,chen2017photographic,9052944}, where feature activation patterns, or deep features, derived from pre-trained auxiliary networks are used to optimize the perceptual quality of output image. Several computational models have been proposed to approximate human perception of audio quality, such as Perceptual Evaluation of Audio Quality (PEAQ) \cite{thiede2000peaq},  Perceptual Evaluation of Speech Quality (PESQ) \cite{rix2001perceptual}, and  Perceptual Evaluation of Audio methods for Source Separation (PEASS) \cite{emiya2010peass}. However, such models are not differentiable, hence cannot be directly employed during TTS training. We believe that utterance level perceptual loss based on deep features that reflects global speech style would be useful to improve overall speech quality.

We are motivated to study a novel training strategy for TTS systems, that learns to associate prosody styles with input text implicitly. We would like to avoid the use of prosody annotations. We don't attempt to model prosody explicitly either, but rather learn the association between prosody styles and input text using existing neural TTS system, such as Tacotron. As the training strategy is only involved during training, it doesn't change the run-time inference process for neural TTS system. At run-time, we don't require any reference signal nor manual selection of prosody style.  

 The main contributions of this paper include: 1) we propose a novel training strategy for Tacotron TTS that improves utterance level expressiveness of speech; 
2) we propose to supervise the training of Tacotron with a fully differentiable perceptual loss, which is derived from a pre-trained auxiliary network, in addition to frame reconstruction loss; and 3) we successfully implement a system that doesn't require any reference speech nor manual selection of prosody style at run-time.
To our best knowledge, this is the first study to incorporate perceptual loss into Tacotron training for improved expressiveness.

This paper is organized as follows: In Section II, we present the research background and related work to motivate our study. In Section III, we propose a novel training strategy for TTS system with frame and style reconstruction loss. In Section IV, we report the subjective and objective evaluations. Section V concludes the discussion.


\section{Background and Related Work}

This work is built on several previous studies on neural TTS, prosody modeling, perceptual loss, and speech emotion recognition. Here we briefly summarize the related previous work to set the stage for our study, and to place our novel contributions in a proper context.

\vspace{-2mm}
\subsection{Tacotron2-based TTS}
\label{sec:baseline}

\begin{figure}[t]
    \centering
    \vspace{2mm}
    \centerline{\includegraphics[width=78mm]{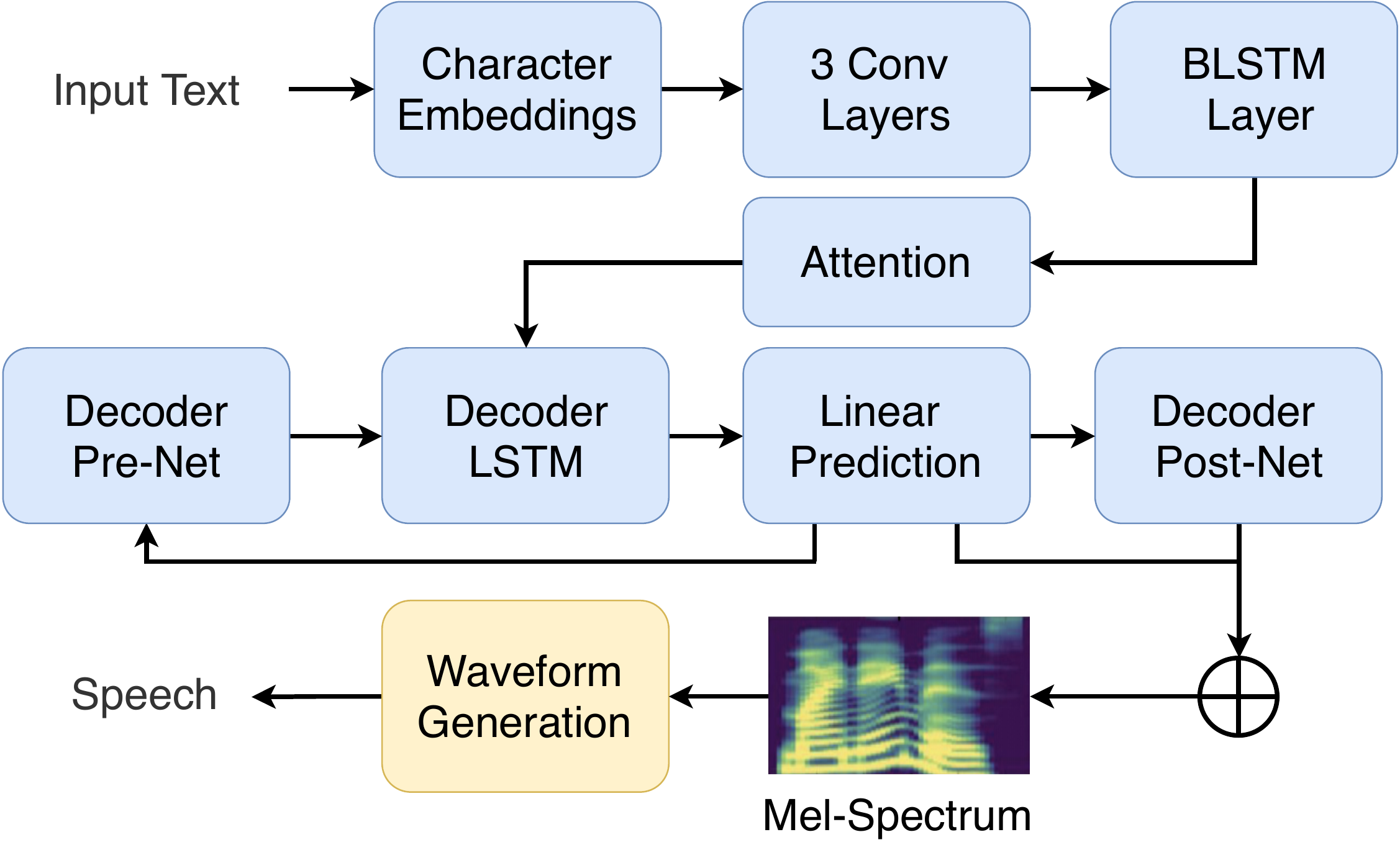}}
   \vspace{2mm}
    \caption{Block diagram of Tacotron2-based TTS reference baseline \cite{shen2018natural}.}
    \label{fig:baseline}
\end{figure}

In this paper, we adopt the Tacotron2-based  \cite{shen2018natural} TTS model as a reference baseline, which is also referred to as \textit{Tacotron} baseline for brevity. 


The overall architecture of the reference baseline includes encoder, attention-based decoder and waveform generation module \cite{griffin1984signal,kalchbrenner2018efficient,oord2016wavenet} as illustrated in Fig. \ref{fig:baseline}. The encoder consists of two components, a convolutional neural network (CNN) module~\cite{krizhevsky2012imagenet, ak2018learning} that has 3 convolutional layers, and a bidirectional LSTM (BLSTM)~\cite{emir2019semantically} layer. The decoder consists of four components: a 2-layer pre-net, 2 LSTM layers, a linear projection layer and a 5-convolution-layer post-net. The decoder is a standard autoregressive recurrent neural network that generates  mel-spectrum features and stop tokens frame by frame. 
There are two common techniques to generate the audio waveform from mel-spectrum features. One is the Griffin Lim \cite{griffin1984signal} algorithm, another is via a neural vocoder~\cite{shen2018natural,oord2016wavenet,kalchbrenner2018efficient, sisman2020overview}.

Just like other TTS systems, {Tacotron}~ \cite{wang2017tacotron,shen2018natural} TTS system predicts  mel-spectrum features from  input sequence of characters by minimizing a frame level reconstruction loss. Such frame level objective function focuses on the distance between spectral features. It does not seek to optimize the perceptual quality at utterance level. 
To improve the suprasegmental expressiveness, there have been studies~\cite{yasuda2019investigation,Stanton2018Predicting,sun2020generating} on latent prosody representations, that make possible prosody styling in Tacotron TTS framework. However, most of the studies rely on the style tokens mechanism to explicitly model the prosody. Simply speaking, they build a Tacotron TTS system that synthesizes speech, and learns the global style tokens (GST) at the same time. At run-time inference, they apply the style tokens to control the expressive effect~\cite{skerry2018towards,wang2018style}, that is referred to as the GST-Tacotron paradigm.

In this paper, we advocate a new way of addressing the expressiveness issue by integrating a perceptual quality motivated objective function into the training process, in addition to the frame level reconstruction loss function. We no longer require any dedicated prosody control mechanism during run-time inference, such as style tokens in Tacotron system.

\vspace{-2mm}
\subsection{Prosody Modeling in TTS}
 
Prosody conveys linguistic, para-linguistic and various types of non-linguistic information, such as speaker identity, intention, attitude and mood\cite{f0_cwt_dct, Ann}. It is inherently supra-segmental \cite{robertt, tokuda2013speech} due to the fact that prosody patterns cannot be derived solely from short-time segments \cite{prosody3}. Prosody is hierarchical in nature \cite{berrak2,prosody3,prosody4, WuProsody} and affected by long-term dependencies at different levels such as word, phrase and utterance level \cite{Sanchez2014}. Studies on hierarchical modeling of F0 in speech synthesis~\cite{Vainio2013, tokuda2013speech, Suni2013} suggest that utterance-level prosody modeling is more effective. Similar studies, such as continuous wavelet transform, can be found in many speech synthesis related applications \cite{Ming2016a, Sanchez2014, Luo2017a, Luo2017, Ming2016}. In this paper, we will study a novel technique to observe utterance-level prosody quality during Tacotron training to achieve expressive synthesis.

The early studies of modeling speaking styles are carried out on Hidden Markov Models (HMM)~\cite{tokuda2002hmm, yamagishi2003modeling}, where we can synthesize speech with an intermediate speaking style between two speakers through model interpolation \cite{tachibana2004hmm}. To improve the HMM-based TTS model, there have been studies to incorporate  unsupervised expression cluster information during training \cite{eyben2012unsupervised}. Deep learning opens up many possibilities for expressive speech synthesis, where  speaker, gender, and age codes can be used as control vectors to change TTS output in different ways~\cite{luong2017adapting}. The style tokens, or prosody embeddings, represent one type of such control vectors, that is derived from a representation learning network. The success of prosody embedding motivates us to further develop the idea.

Tacotron TTS framework has achieved remarkable performance in terms of spectral feature generation. With a large training corpus, it may be able to generate natural prosody and expression by remembering the training data using a large number of network parameters. However, its training process doesn't aim to optimize the system for expressive prosody rendering. As a result, Tacotron TTS system tends to generate speech outputs that represent model average, rather than the intended prosody.

The idea of global style tokens~\cite{wang2018style,Stanton2018Predicting} represents a success in controlling  prosody style of Tacotron output. Style tokens learn to represent high level styles, such as speaker style, pitch range, and speaking rate across a collection of utterances or a speech database. We argue that they neither necessarily represent the useful styles to describe the continuum of prosodic expressions~\cite{kenter2019chive}, nor provide the dynamic and differential prosodic details with the right level of granularity at utterance level. Sun et al. \cite{sun2020fully,sun2020generating} study a way to include a hierarchical, fine-grained prosody representation, that represents the recent attempts to address the problems in GST-Tacotron paradigm.

We would like to address three issues in the existing prosody modeling in Tacotron framework, 1) lack of prosodic supervision during  training; 2) limitation of explicit prosody modeling, such as style tokens, in describing the continuum of prosodic expressions; 3)  lack of dynamic and differential prosody at utterance level.

\vspace{-4mm}
\subsection{Perceptual Loss for Style Reconstruction}

It is noted that frame-level reconstruction loss, denoted as \textit{frame reconstruction loss} in short, is not always consistent with human perception because it doesn't take into account human sensitivities to temporal and spectral information, such as prosody and temporal structure of the utterance. For example, if one repeatedly asks the same question two times, despite the perceptual similarity of two utterances, they would be very different as measured by frame-level losses.  

Perceptual loss refers to the training loss derived from a pre-trained auxiliary network~\cite{johnson2016perceptual}. The auxiliary network is usually trained on a different task that provides perceptual quality evaluation of an input at a higher level than a speech frame. The intermediate feature representations, generated by the auxiliary network in form of hidden layer activations, are usually referred to as deep features. They are used as the high level abstraction to measure the training loss between reconstructed signals and reference signals. Such training loss is also called deep feature loss~\cite{9053110, 9054578}.

In speech enhancement, perceptual loss has been used successfully in end-to-end speech denoising pipeline, with an auxiliary network pre-trained on audio classification task \cite{germain2019speech}. 
Kataria et al. \cite{9053110} propose to use perceptual loss which optimizes the enhancement network with an auxiliary network pre-trained on speaker recognition task. In voice conversion, Lo et al.~\cite{lo2019mosnet} propose deep learning-based assessment models to predict human ratings of converted speech.
Lee~\cite{lee2020voice} propose a perceptually meaningful criterion where human auditory system was taken into consideration in measuring the distances between the converted speech and the reference.  

In speech synthesis, Oord et al. propose  to train a WaveNet-like classifier with perceptual loss for phone recognition~\cite{pmlr-v80-oord18a}. As the classifier extracts high-level features that are relevant for phone recognition, this loss term supervises the training of WaveNet to look after temporal dynamics, and penalize bad pronunciations. Cai et al.~\cite{cai2020speaker} study to use a pre-trained speaker embedding network to provide feedback constraint, that serves as the perceptual loss for the training of  a multi-speaker TTS system.

In the context of prosody modeling, the perceptual loss in the above studies can be generally described as \textit{style reconstruction  loss}~\cite{johnson2016perceptual}. Following the same principle, we would like to propose a novel auxiliary network, that is pre-trained on a speech emotion recognition (SER) task, to extract high level prosody representations. By comparing prosody representations in a continuous space, we measure perceptual loss between two utterances. While perceptual loss is not new in speech reconstruction, the idea of using a pre-trained emotion recognition network for perceptual loss is a novel attempt in speech synthesis.

\vspace{-4mm}
\subsection{Deep Features for Perceptual Loss}

Now the question is which deep features could be suitable for measuring perceptual loss. We benefit from the prior work in prosody modeling.
Prosody embedding in Tacotron is a type of feature learning, that learns the representation for prediction or classification tasks. With deep learning algorithms, automatic feature learning can be achieved in either supervised, such as multilayer perceptron~\cite{bottleneckSER}, or unsupervised manner, such as variational autoencoder \cite{kingma2014auto}. Deep features
are usually more
generalizable, and easier to manage than hand-crafted or manually designed features~\cite{Zhong2016AnOO}. There have been studies on representation learning for prosody patterns, such as speech emotion~ \cite{vaefeaturelearning}, and speech styles\cite{wang2018style}.

\textcolor{black}{
Affective prosody refers to the expression of emotion in speech  \cite{Zhang2018SpeechER,10.1109/TASLP.2019.2898816}. It is prominently exhibited in emotion speech database. Therefore, the studies in speech emotion recognition provide valuable insights into prosodic modeling. Emotion are usually characterized by discrete categories, such  as  happy,  angry, and sad,  and continuous attributes, such as activation, valence and dominance~\cite{murray1993toward,pierre2003production}. Recent studies show that latent representations of deep neural networks also characterize well emotion in a continuous space~\cite{bottleneckSER}. } 

 \textcolor{black}{
There have been studies to leverage emotion speech modeling for expressive TTS \cite{eyben2012unsupervised,skerry2018towards,wu2019end,gao2020interactive,um2020emotional}. Eyben et al. \cite{eyben2012unsupervised} incorporate  unsupervised expression cluster information into an HMM-based TTS system. Skerry-Ryan et al.~\cite{skerry2018towards} study learning prosody representation from animated and emotive storytelling speech corpus. Wu et al. \cite{wu2019end} propose a semi-supervised training of Tacotron TTS framework for emotional speech synthesis, where style tokens are defined to represent emotion categories.  
Gao et al. \cite{gao2020interactive} propose to use an emotion recognizer to extract the style embedding for speech style transfer.  Um et al. \cite{um2020emotional} study a technique to apply style embedding to Tacotron system to generate emotional speech, and to control the intensity of emotional expressiveness.  }

\textcolor{black}{
All the studies point to the fact that emotion-related deep features serve as the excellent descriptors of speech prosody and speech styles. In this paper, instead of using the style tokens to control the TTS outputs, we would like to study how to use deep style features to measure perceptual loss for the training of neural TTS system in general.}

\begin{figure}[t]
    \centering
    \centerline{\includegraphics[width=0.95\linewidth]{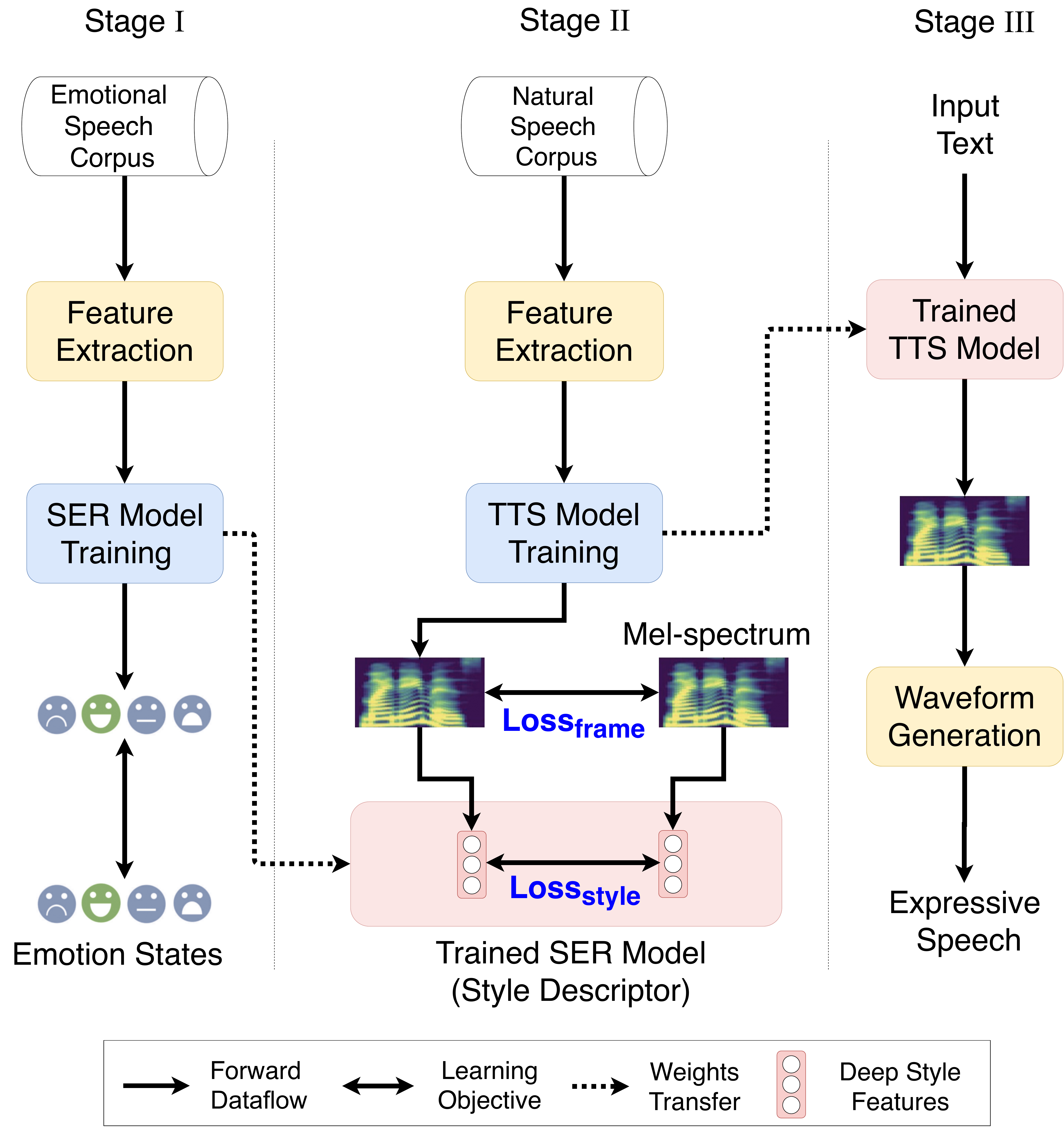}}
    \caption{ Overall framework of a \textit{Tacotron-PL} system in three stages: Stage I for training of style descriptor; Stage II for training of \textit{Tacotron-PL}; Stage III for run-time inference. 
    }
    \label{fig:overallsummary}
\end{figure}

\section{Tacotron with Frame and Style Reconstruction Loss}
\label{sec:model}

We propose a novel training strategy for Tacotron with both frame and style reconstruction loss. As the style reconstruction loss is formulated as a perceptual loss (PL)~\cite{johnson2016perceptual}, the proposed frame and style training strategy is called \textit{Tacotron-PL} in short. It seeks to optimize both frame-level spectral loss, that is \textit{frame reconstruction loss}, as well as utterance-level style loss, that is \textit{style reconstruction loss}, at the same time. 

The overall framework is illustrated in Fig. \ref{fig:overallsummary}, that has three stages: 1) training of style descriptor, 2) the proposed frame and style training for \textit{Tacotron-PL} model, and 3) run-time inference. In Stage I, we train an auxiliary network to serve as the style descriptor for input speech utterances. In Stage II,  the proposed frame and style training strategy is implemented to associate input text with  acoustic features, as well as  prosody style of natural speech, that is assisted by the style descriptor obtained from Stage I.  
In Stage III, the {\textit{Tacotron-PL}} system takes input text and generates expressive speech in the same way as a standard Tacotron does. Unlike other Tacotron variants \cite{wang2018style}, \textit{Tacotron-PL} doesn't require any add-on module or process for run-time inference.  

As discussed in Section II-A,  traditional Tacotron architecture contains a text encoder and an attention-based decoder. We first encode input character embedding into hidden state, from which the decoder generates mel-spectrum features. During training, we adopt a frame-level mel-spectrum loss as in \cite{shen2018natural}, which is a $L_{2}$ loss between the synthesized mel-spectrum $\hat{\textbf{Y}} = \{ \hat{\textbf{y}}_{1}, ... \hat{\textbf{y}}_{t}, ...\hat{\textbf{y}}_{T} \} $ and target mel-spectrum $\textbf{Y} = \{ \textbf{y}_{1}, ... \textbf{y}_{t}, ...\textbf{y}_{T} \} $. We have $Loss_{frame}$ as follows,

\begin{equation}
\label{eq:loss1}
Loss_{frame}({\textbf{Y}}, {\hat{\textbf{Y}}}) = {\sum_{t=1}^{{T}}  L_{2} (\textbf{y}_{t},\hat{\textbf{y}}_{t})}  
\end{equation}
which is designed to minimize frame level distortion. As it doesn't guarantee utterance level similarity concerning speech expressions, such as speech prosody and speech styles. We will study a new loss function $Loss_{style}$ next, that measures the utterance-level style reconstruction loss.

\begin{figure*}[t]
    \centering
    \centerline{\includegraphics[width=184mm]{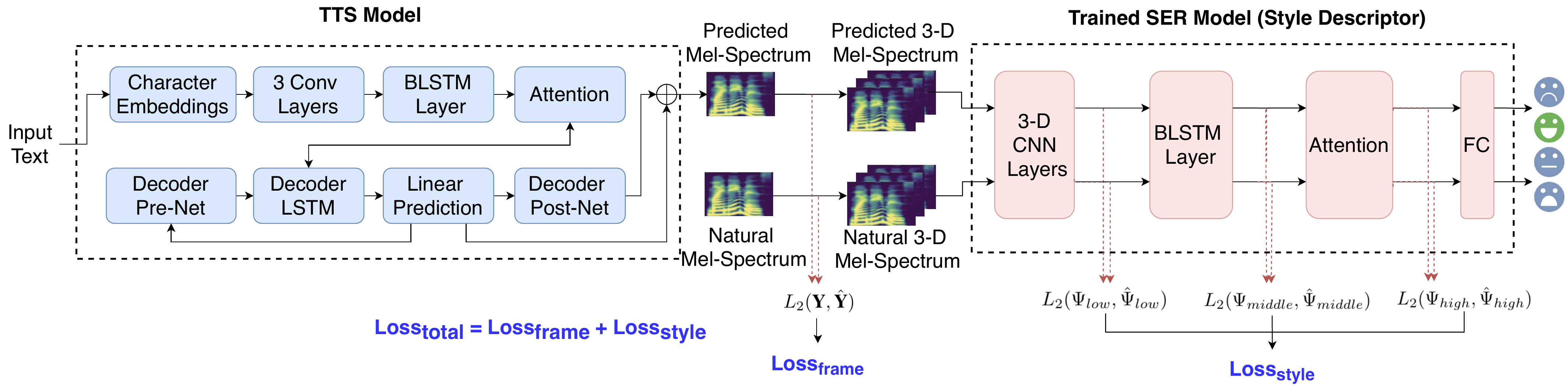}}
    \caption{Block diagram of the proposed training strategy, \textit{Tacotron-PL}. A speech emotion recognition (SER) model is trained separately to serve as an auxiliary model to extract deep style features. A \textit{style reconstruction loss}, $Loss_{style}$, is computed between the deep style features of the generated  and reference speech at utterance-level. }
    \label{fig:model}
\end{figure*}

\subsection{{Stage I:} Training of Style Descriptor}
\label{subsec:pretraining}
 
One of the great difficulties of prosody modeling  is the lack of reference samples. In linguistics,  we usually describe prosody styles qualitatively. However, precise annotation of speech prosody is not straightforward. One of the ways to describe a prosody style is to show by example. The idea of style token~\cite{wang2018style} shows a way to compare two prosody styles quantitatively using deep features.

Manual prosodic annotations of recorded speech~\cite{Silverman1992TOBIAS} provide quantifiable prosodic labels that allow us to associate speech styles with actual acoustic features. Prosody labeling schemes often attempt to describe prosodic phenomena, such as the supra-segmental features of intonation, stress, rhythm and speech rate, in discrete categories. Categorical labels of speech emotion~\cite{busso2008iemocap} also seek to achieve a similar goal. The prosody labeling schemes serve as a type of style descriptor. With deep neural network, one is able to learn the feature representation of the data at different level of abstraction in a continuous space~\cite{goodfellow2016deeplearning}. As speech styles naturally spread over a continuum rather than forced-fitting into a finite set of categorical labels, we believe that deep neural network learned from animated and emotive speech serves as a more suitable style descriptor.

We propose to use a speech emotion recognizer  (SER)~\cite{Zhang2018SpeechER,10.1109/TASLP.2019.2898816} as a style descriptor $F(\cdot)$, which extracts deep style features $\Psi$ from an utterance $\textbf{Y}$, or    $\Psi=F(\textbf{Y})$.
We use neuronal activations of hidden units in a deep neural network as the deep style features to represent high level prosodic abstraction at utterance level. In practice, we first train an SER network with highly animated and emotive speech with supervised learning. 
\textcolor{black}{We then derive deep style features from a small intermediate layer. As the intermediate layer is small relative to the size of the other layers, it  creates  a  constriction  in  the  network  that  forces  the  information  pertinent  to  emotion classification into  a  low  dimensional prosody representation~\cite{bottlenectDong}. Such low dimensional prosody representation is expected to describe the prosody style of speech signals as the SER network relies on the prosody representation for accurate emotion classification.}


We follow the SER implementation in ~\cite{chen20183, zhou2020seen} as illustrated in Fig. \ref{fig:model}, that forms part of Fig. \ref{fig:overallsummary}.  The SER network includes 1) a three-dimensional (3-D) CNN layer; 2) a BLSTM layer~\cite{Greff2017LSTM}; 3) an attention layer; and 4) a fully connected (FC) layer. The 3-D CNN~\cite{chen20183} first extracts a latent representation from  mel-spectrum, its delta and delta-delta values from input utterance, converting the input utterance of variable length into a fixed size latent representation, denoted as deep features sequence $\Psi_{low}$, that reflects the semantics of emotion. The BLSTM summarizes the temporal information of $\Psi_{low}$ into another latent representation $\Psi_{middle}$. Finally, the attention layer assigns weights to $\Psi_{middle}$ and generates $\Psi_{high}$ for emotion prediction.  
 
The question is which of the latent representations, $\Psi_{low}$, $\Psi_{middle}$, and $\Psi_{high}$, is suitable to be the deep style features. To validate the descriptiveness of deep style features, we perform an analysis on LJ-Speech corpus~\cite{ljspeech17}. Specifically, we randomly select five utterances from each of the six style groups from the database, each group having a distinctive speech style, namely, 1) Short question; 2) Long question; 3) Short answer; 4) Short statement; 5) Long statement and 6) Digit string. The complete list of utterances can be found at Table \ref{tab:textgroup} in Appendix A. 

\begin{figure}[!t]
\centering
\includegraphics[width=50mm]{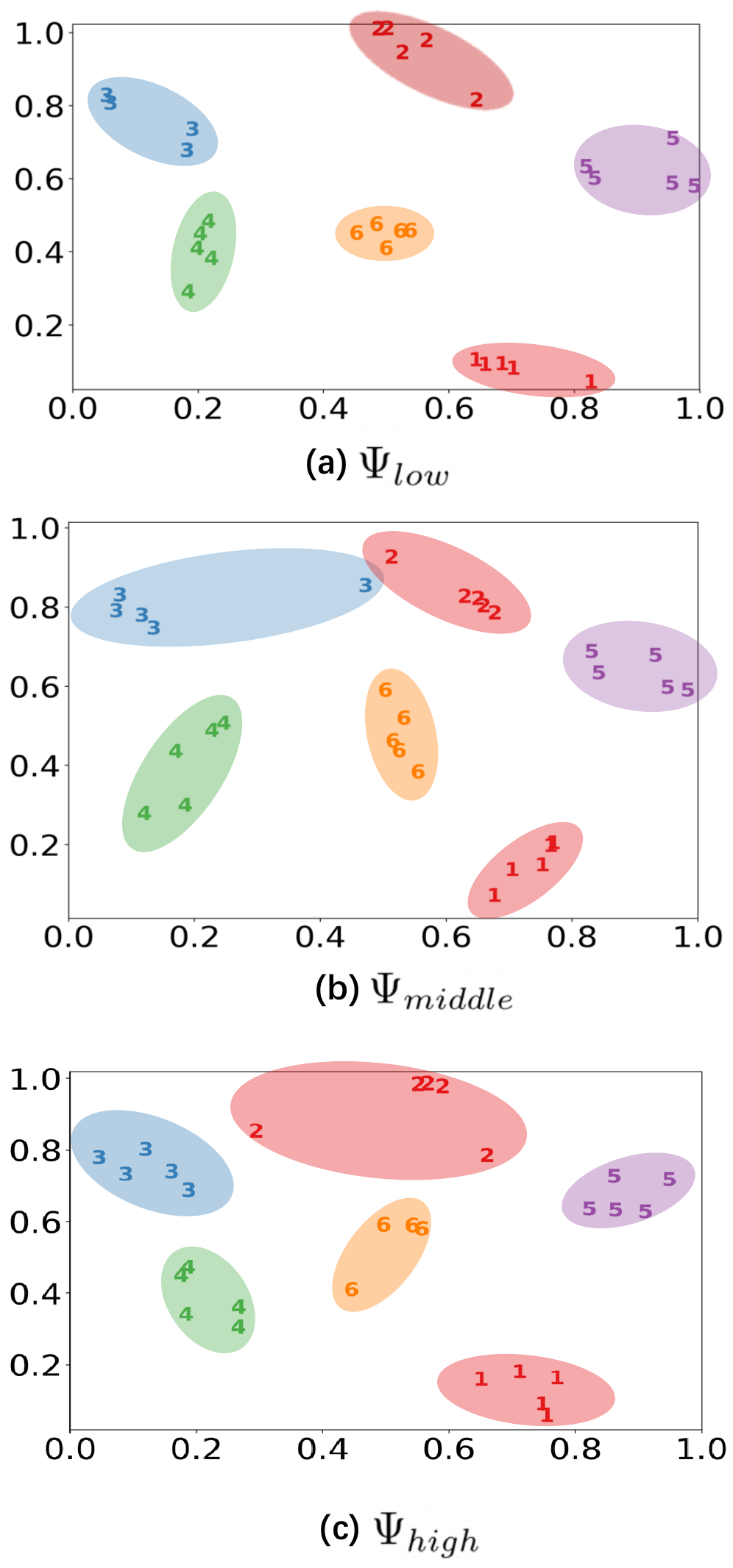}
\caption{t-SNE plot of the distributions of deep style features $\Psi_{low}$, $\Psi_{middle}$ and $\Psi_{high}$ for six groups of utterances in LJ-Speech corpus. The list of utterances can be found at Table V in Appendix A.}
\label{fig:pe}
\end{figure}

We visualize the $\Psi_{low}$, $\Psi_{middle}$ and $\Psi_{high}$ of utterances using the t-SNE algorithm in a two dimensional plane~ \cite{maaten2008visualizing}, as shown in Fig. \ref{fig:pe}.
Please note that the distributions of digits 1 to 6 represent those of groups 1 to 6 in the two dimensional space. As illustrated in Table V, the utterances within the same group form a cluster, while the utterances between groups distance from one another. To visualize, we color the clusters to highlight their distributions.
It is observed that $\Psi_{low}$, $\Psi_{middle}$ and $\Psi_{high}$ of utterances form clear style groups in terms of feature distributions, that correspond to the six different utterance styles summarized in Table V. Furthermore, it is clear that Fig. \ref{fig:pe}(a) shows a better clustering than Fig. \ref{fig:pe}(b) and Fig. \ref{fig:pe}(c). We will further compare the performance of different deep style features through TTS experiments in Section IV.

\subsection{{Stage II}: \textit{Tacotron-PL} Training}

During the training of \textit{Tacotron-PL}, the SER-based style descriptor $F(\cdot)$ is used to extract the  deep  style features $\Psi$. We define a style reconstruction loss that compares the  prosody style between the reference speech ${\textbf{Y}}$ and the generated speech ${\hat{\textbf{Y}}}$ .

\begin{equation}
Loss_{style}({\textbf{Y}}, {\hat{\textbf{Y}}})  =  L_{2} ( \Psi,\hat{\Psi})
\end{equation}
where  ${\Psi}=F({\textbf{Y}})$ and $\hat{\Psi}=F(\hat{\textbf{Y}})$. As illustrated in Fig. \ref{fig:model}, the proposed training strategy involves two loss functions: 1) $Loss_{frame}$ that minimizes the loss between synthesized and original mel-spectrum at frame level; and 2) $Loss_{style}$ that minimizes the style differences between the synthesized and reference speeches at utterance level.
 \begin{align}
    Loss_{total}({\textbf{Y}}, {\hat{\textbf{Y}}}) =Loss_{frame}({\textbf{Y}}, {\hat{\textbf{Y}}}) + Loss_{style}({\textbf{Y}}, {\hat{\textbf{Y}}})
\end{align}
where $Loss_{frame}$ is also the loss function of a traditional Tacotron~\cite{shen2018natural} system. 

Style reconstruction loss can be seen as perceptual quality feedback at utterance level to supervise the training of prosody style. 
All parameters in the TTS model are updated with the gradients of the total loss through back-propagation. We expect that mel-spectrum generation will learn from local and global viewpoint through the frame and style reconstruction loss.

\subsection{{Stage III:} Run-time Inference}
The inference stage follows exactly the same Tacotron workflow, that only involves the TTS Model in Fig. \ref{fig:model}. The difference between \textit{Tacotron-PL} and other global style tokens variation of Tacotron is that \textit{Tacotron-PL} encodes  prosody styling inside the standard Tacotron architecture. It doesn't require any add-on module. 

At run-time, the Tacotron architecture takes text as input and generate expressive mel-spectrum features as output, that is followed by Griffin-Lim algorithm \cite{griffin1984signal} and WaveRNN vocoder \cite{kalchbrenner2018efficient} in this paper to generates audio signals.

\section{Experiments}
\label{sec:exp}

\textcolor{black}{We train a SER as the style descriptor on IEMOCAP dataset \cite{busso2008iemocap}, which consists of five sessions.
The dataset contains a total of 10,039 utterances, with an average duration of 4.5 seconds at a sampling rate of 16 kHz. We only use a subset of the improvised data with four emotional categories, namely, happy, angry, sad, and neutral, which are recorded in the hypothetical scenarios designed to elicit specific  types of emotions.} 

With the style descriptor, we further train a Tacotron system on LJ-Speech database \cite{ljspeech17}, which consists of 13,100 short clips with a total of nearly 24 hours of speech from one single speaker reading 7 non-fiction books. The speech samples are available from the demo link \footnote{
\textbf{Speech Samples:} \textcolor{magenta}{\href{https://ttslr.github.io/Expressive-TTS-Training-with-Frame-and-Style-Reconstruction-Loss/}{https://ttslr.github.io/Expressive-TTS-Training-with-Frame-and-Style-Reconstruction-Loss/}} }.

\begin{figure}[t]
    \centering
    \centerline{\includegraphics[width=80mm]{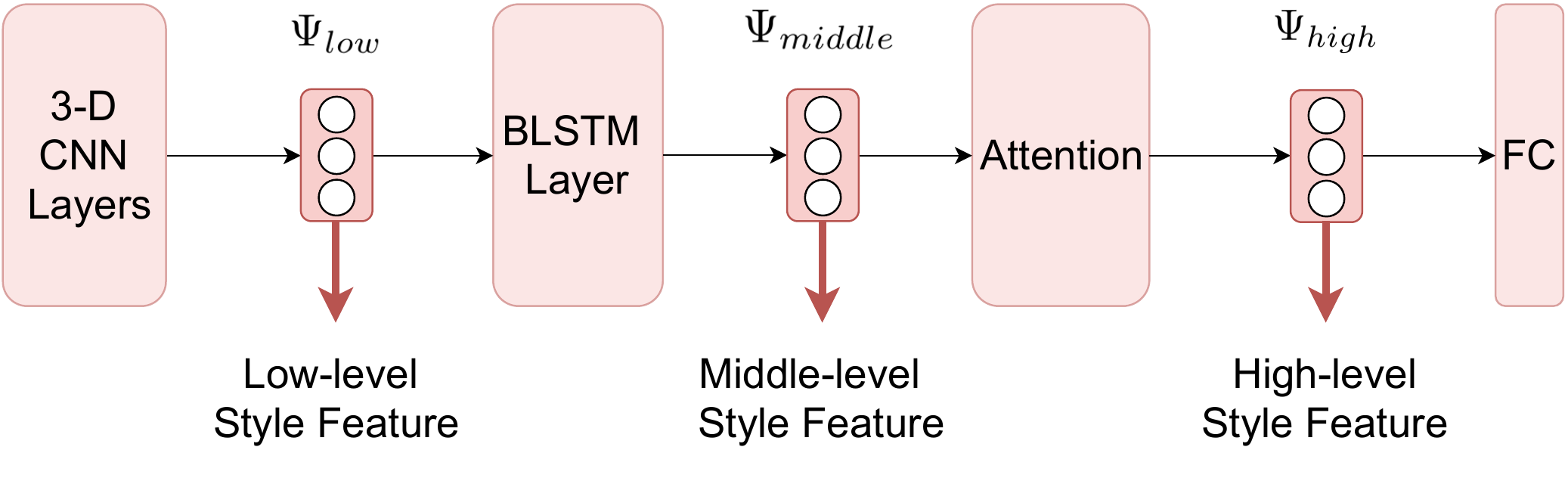}}
    \caption{Three level (low, middle and high) of deep style features extracted from SER-based style descriptors for computing style construction loss. }
    \label{fig:systems}
\end{figure}

\subsection{Comparative Study}

We develop five Tacotron-based TTS systems for a comparative study, that includes the Tacotron baseline, 
and four variants of Tacotron with the proposed training strategy, \textit{Tacotron-PL}.

To study the effect of different style descriptors, we compare the use of four deep style features, which includes three single features and a combination of them, in $Loss_{style}$, as illustrated in Fig. \ref{fig:systems}, and summarized as follows:

\begin{itemize}
    \item \textit{Tacotron}: Tacotron \cite{shen2018natural} trained with $Loss_{frame}$ as in Eq. (1), that doesn't explicitly model speech style. 
    \item \textit{Tacotron-PL(L)}: \textit{Tacotron-PL}  which uses $\Psi_{low}$ in $Loss_{style}$.
    \item \textit{Tacotron-PL(M)}: \textit{Tacotron-PL}  which uses $\Psi_{middle}$ in $Loss_{style}$.
    \item \textit{Tacotron-PL(H)}: \textit{Tacotron-PL}  which uses $\Psi_{high}$ in $Loss_{style}$.
    \item \textit{Tacotron-PL(LMH)}: \textit{Tacotron-PL}  which uses $\{ \Psi_{low}, \Psi_{middle},  \Psi_{high} \}$ in $Loss_{style}$.
\end{itemize}

\subsection{Experimental Setup}
\label{subsec:expsetup}
\textcolor{black}{For SER training, we first split the speech signals into segments of 3 seconds as in~\cite{chen20183}. We then extract 40-channel mel-spectrum features with a frame size of 50ms and 12.5ms frame shift. The first convolution layer has 128 feature maps, while the remaining convolution layers have 256 feature maps. The filter size for all convolution layers is 5$\times$3, with 5 along the time axis, and 3 along the frequency axis, and the pooling size for the max pooling layer is 2$\times$2.  
We add a linear layer with 200 output units after 3-D CNN for dimension reduction.} 

\textcolor{black}{In this way, the 3-D CNN extracts a fixed size of latent representation with $150 \times 200$  dimension from the input utterance, that we use as the deep style features $\Psi_{low}=F_{low}(\cdot)$ to represent a temporal sequence of $150$ segment, each having an embedding of $200$ elements. 
As each direction of BLSTM layer contains $128$ cells, in two directions, we obtain $256$ output activations for each input segment, that are further mapped to $200$ output units via a linear layer.  BLSTM summarizes the temporal information of $\Psi_{low}$ into another fixed size latent representation $\Psi_{middle}=F_{middle}(\cdot)$ of $150 \times 200$ dimension. The attention layer assigns the weights to $\Psi_{middle}$ and generate a new latent representation $\Psi_{high}=F_{high}(\cdot)$. All latent representation $\Psi_{low}$, $\Psi_{middle}$, $\Psi_{high}$ have the same dimension.} 

\textcolor{black}{The fully connected layer contains 64 output units. Batch normalization \cite{ioffe2015batch} is applied to the fully connected layer to accelerate training and improve the generalization performance.
The parameters of the SER model were optimized by minimizing the cross-entropy objective function, with a minibatch of 40 samples, using the Adam optimizer with Nestorov momentum. The initial learning rate is set to $10^{-4}$ and the momentum is set to 0.9. In this way, we obtain a SER style descriptor that is reported with an average classification accuracy of 73.2\% for all emotions on the test set.}

The SER-based style descriptor is used to extract deep style features for the computing of $Loss_{style}$. 
For TTS training, the encoder takes a 256-dimensions character sequence as input and the decoder generates the 40-channel mel-spectrum.
The training utterances from LJ-Speech database are of variable length. Mel-spectrum features are also extracted with a frame size of 50ms and 12.5ms frame shift. They are normalized to zero-mean and unit-variance to serve as the reference target. The decoder predicts only one non-overlapping output frame at each decoding step. We use the Adam optimizer with $\beta_1$ = 0.9, $\beta_2$ = 0.999 and a learning rate of $10^{-3}$ exponentially decaying to $10^{-5}$ starting at 50k iterations. We also apply $L_{2}$ regularization with weight $10^{-6}$. All models are trained with a batch size of 32 and 150k steps.

\begin{table}[t]
\centering
\caption {The MCD, RMSE and FD results of different systems.}
\label{tab:mcd}
\begingroup
\begin{tabular}{p{2.8cm}p{1.4cm}p{1.5cm}p{1.4cm}}
\toprule
 \textbf{System}&  \textbf{MCD} [dB] & \textbf{RMSE} [Hz] & \textbf{FD} [frame]  \\
\hline 
Tacotron & 7.01 & 1.53 & 15.59 
\\
Tacotron-PL(L) & \textbf{6.37}  & \textbf{0.94 } & \textbf{13.96}
\\
Tacotron-PL(M) & 6.70  & 1.21  & 14.20 
\\
Tacotron-PL(H) & 6.88 & 1.42  & 15.41 
\\
Tacotron-PL(LMH) & 6.62 & 1.16 & 14.15 
\\
\bottomrule
\end{tabular}
\endgroup
\end{table}

\begin{figure}[t]
    \centering
    \centerline{\includegraphics[width=0.46\textwidth]{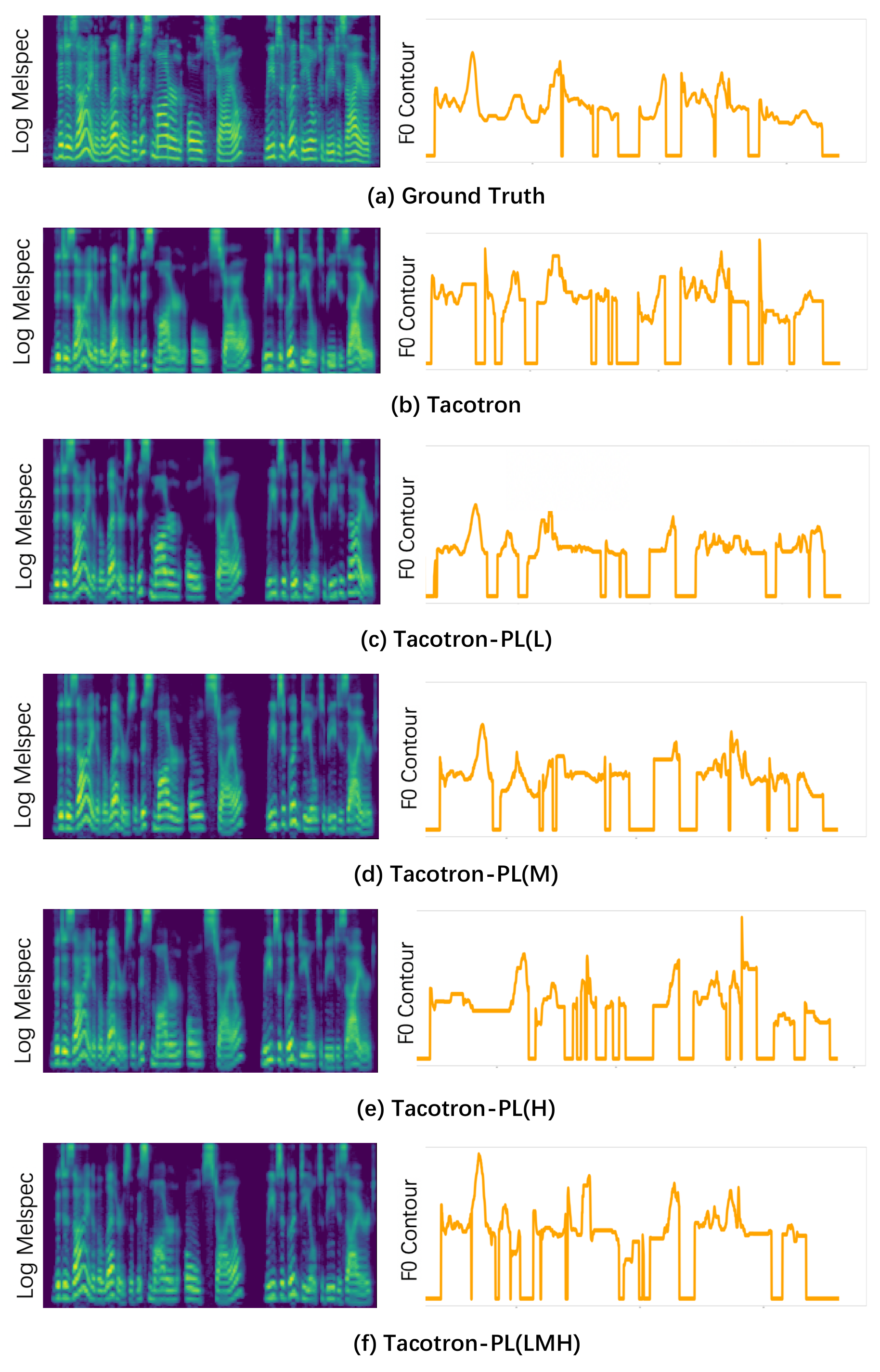}}
    \vspace{-2mm}
    \caption{Spectrogram (left) and F0 contour (right) of an utterance ``\textit{The design of the letters of this modern `old style' leaves a good deal to be desired.}'' from LJ-Speech database between the reference natural speech, labelled as Ground Truth,  and five Tacotron systems. It is observed that \textit{Tacotron-PL} models produce finer spectral details, prosodic phrasing and F0 contour that are closer to those of the reference than \textit{Tacotron} baseline.}
    \label{fig:melf0}
\end{figure}

\subsection{Objective Evaluation}
\label{subsec:obj}
We conduct objective evaluation experiments to compare  the systems in a comparative study. The results are summarized in Table I. 

\subsubsection{Performance Evaluation Metrics}

Mel-cepstral distortion (MCD) \cite{kubichek1993mel} is used to measure the spectral distance between the synthesized and reference mel-spectrum features that is known to correlate well with human perception \cite{kubichek1993mel}. MCD is calculated as:
\begin{equation}
{\rm MCD} =\frac{10 \sqrt{2}}{\ln 10} \frac{1}{N}  \sqrt{  \sum_{k=1}^{N}\left({y}_{t,k}-\hat{{y}}_{t,k}\right)^{2}}
\end{equation}
where $N$ represents the dimension of the mel-spectrum, ${y}_{t,k}$ denotes the $k^{th}$ mel-spectrum component in $t^{th}$ frame for the reference target mel-spectrum, and $\hat{{y}}_{t,k}$ for the synthesized mel-spectrum. Lower MCD value indicates smaller distortion.

We use Root Mean Squared Error (RMSE) as the evaluation metrics for F0 modeling, that is calculated as:
\begin{equation}
{\rm R M S E}=\sqrt{\frac{1}{T} \sum_{t=1}^{T}\left( {\rm F 0_{t}} - \widehat{\rm F 0}_{t} \right)^{2}}
\end{equation}
where ${\rm F 0_{t}}$ and $\widehat{\rm F 0_{t}}$ denote the  reference and synthesized  F0 at $t^{th}$ frame. We note that lower RMSE value suggests that the two F0 contours are more similar.

Moreover, we propose to use frame disturbance, denoted as FD, to calculate the deviation in the dynamic time warping (DTW) alignment path \cite{8300542,jusoh2015investigation,gupta2017perceptual}. 
FD is calculated as:
\begin{equation}
{\rm FD}= \sqrt{\frac{1}{T}\sum_{t=1}^{T}\left( { \rm  a_{t,x}} - {\rm  a_{t,y}} \right)^{2}}
\end{equation}
where ${\rm a_{t,x}}$ and $ {\rm a_{t,y}}$ denote the x-coordinate and the y-coordinate of the $t^{th}$ frame in the DTW alignment path. As FD represents the duration deviation of the synthesized speech from the target, it is a proxy to show the duration distortion. A larger value indicates poor duration modeling performance and a smaller value indicates otherwise.

\subsubsection{Spectral Modeling}
We observe that all implementations of \textit{Tacotron-PL} model consistently provide lower MCD values than  \textit{Tacotron} baseline, with   \textit{Tacotron-PL(L)}  representing the lowest MCD, as can be seen in Table \ref{tab:mcd}.  
We also visualize the spectrograms of same speech content synthesized by five different models, together with that of the reference natural speech in Fig. \ref{fig:melf0}. A visual inspection of the spectrograms suggests that \textit{Tacotron-PL} models consistently provide finer spectral details than \textit{Tacotron} baseline. 

\begin{figure}[t]
    \centering
    \centerline{\includegraphics[width=0.46\textwidth]{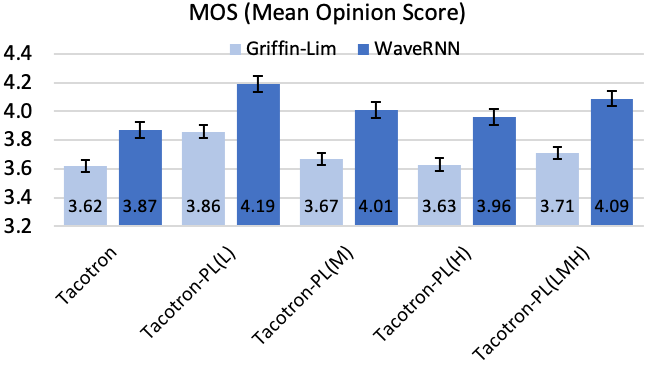}}
    \caption{The mean opinion scores (MOS) of five systems evaluated by 15 listeners, with 95\% confidence intervals computed from the t-test.}
    \label{fig:mos}
\end{figure}

\subsubsection{F0 Modeling}

Fundamental frequency, or F0, is an essential prosodic feature of speech~ \cite{Stanton2018Predicting, sun2020generating}. As there is no guarantee that synthesized speech and reference speech have the same length, we apply DTW~\cite{muller2007dynamic} to align speech pairs and calculate RMSE between the F0 contour of them. The results are reported in Table \ref{tab:mcd}. It is observed that \textit{Tacotron-PL} models consistently generate F0 contours which are closer to reference speech than \textit{Tacotron} baseline. 

We note that both F0 and prosody style contributes to RMSE measurement. To show the effect of various deep style features on the F0 contours, we also plot the F0 contours of the utterances in Fig. \ref{fig:melf0}. A visual inspection suggests that the \textit{Tacotron-PL} models benefit from the perceptual loss training, and produce F0 contour with a  better fit to that of the reference speech, with \textit{Tacotron-PL(L)} producing the best fit (see Fig. \ref{fig:melf0}(c)).

\subsubsection{Duration Modeling}

Frame disturbance is a proxy to the duration difference~\cite{gupta2017perceptual} between  synthesized speech and  reference natural speech. We report frame disturbance of five systems in Table \ref{tab:mcd}. 
As shown in Table \ref{tab:mcd},  \textit{Tacotron-PL}  models obtain significantly lower FD value than  \textit{Tacotron}  baseline, with \textit{Tacotron-PL(L)} giving the lowest FD. From Fig. \ref{fig:melf0}, we can also observe that \textit{Tacotron-PL(L)} example clearly provides a better duration prediction than other models.
We can conclude that perceptual loss training with style reconstruction loss helps Tacotron to achieve a more accurate rendering of prosodic patterns.

\begin{table}[]
\centering
\caption{The AB preference test for expressiveness and naturalness evaluation by 15 listeners, with 95\% confidence intervals computed from the t-test.}
\begingroup
\renewcommand{\arraystretch}{1} 
\begin{tabular}{p{4.5cm}p{0.6cm}p{0.6cm}p{0.6cm}<{\centering}p{0.7cm}<{\centering}}
\toprule
\multirow{2}{*}{\textbf{\tabincell{c}{Contrastive pair}
}} & \multicolumn{3}{l}{\textbf{Preference}(\%)} & \multirow{2}{*}{$\bm{p}$\textbf{-value}} \\ \cline{2-4}
 & \textbf{Former} & \textbf{Neutral} & \textbf{Latter} &  \\ \hline
 \multicolumn{5}{l}{\textbf{Expressiveness}}  \\ \hline
 Tacotron vs. Tacotron-PL(L)  & 32.44 & 13.33 & 54.23 &  0.00119\\
Tacotron--PL(LMH) vs. Tacotron-PL(L)  & 37.78  & 11.56  & 50.66 &  0.00124\\  \hline
\multicolumn{5}{l}{\textbf{Naturalness}} \\ \hline
Tacotron vs. Tacotron-PL(L)  & 36.44 & 18.22 & 45.34 &  0.00101\\
Tacotron-PL(LMH) vs. Tacotron-PL(L)  & 39.11 & 15.56  & 45.33 &  0.00096\\\bottomrule
\end{tabular}
\endgroup
\label{tab:ab1}
\end{table}

\begin{table}[t]
\centering
\caption {Best Worst Scaling (BWS) listening experiments that compare {four deep style features} in four \textit{Tacotron-PL} models.}
\begin{tabular}{p{3.4cm}p{2.2cm}p{2.2cm}}
\toprule
 \textbf{System}&  \textbf{Best} (\%) & \textbf{Worst} (\%)  \\
\hline 
Tacotron-PL(L) & \textbf{80}  & \textbf{5}  \\
Tacotron-PL(M) & 8 & 26  \\
Tacotron-PL(H) & 2 & 48    \\
Tacotron-PL(LMH) &10 & 21  \\
\bottomrule
\end{tabular}
\label{tab:bws}
\end{table}

\subsubsection{Deep Style Features} 
We compare {four different deep style features} by evaluating the performance of their use in \textit{Tacotron-PL} models, namely \textit{Tacotron-PL(L)}, \textit{Tacotron-PL(M)}, \textit{Tacotron-PL(H)} and \textit{Tacotron-PL(LMH)}. 

\textcolor{black}{In supervised feature learning, the features that are near the input layer are related to the low level features, while those that are near the output are related to the supervision target, that are the categorical labels of the emotion.  While we expect the style descriptors to capture utterance level prosody style, we don't want the  style reconstruction loss function to  directly relate to emotion categories. Hence, the lower level deep  features, $\Psi_{low}$, as illustrated in Fig. \ref{fig:pe}, would be more appropriate than the higher level deep features, such as $\Psi_{middle}$ and $\Psi_{high}$.}

 \textcolor{black}{
We observe that $\Psi_{low}$ is more descriptive than other deep style features for perceptual loss evaluation, as reported in spectral modeling (MCD), F0 modeling (RMSE), duration modeling (FD) for \textit{Tacotron-PL} experiment in Table \ref{tab:mcd}. The observations confirm our intuition and the analysis in Fig. \ref{fig:pe}. }



\vspace{-2mm}
\subsection{Subjective Evaluation}

We conduct listening experiments to evaluate several aspects of the synthesized speech, and the choice of deep style features for $Loss_{style}$. Griffin-Lim algorithm \cite{griffin1984signal} and neural vocoder are employed to generate the speech waveform. We choose WaveRNN vocoder which follows the same parameter settings as \cite{kalchbrenner2018efficient} since it's the first sequential neural model for real-time audio synthesis \cite{kalchbrenner2018efficient}.

\subsubsection{Voice Quality}
 
Each audio is listened by 15 subjects, each of which listens to 150 synthesized speech samples. 
We first evaluate the voice quality in terms of mean opinion score (MOS) among \textit{Tacotron}, \textit{Tacotron-PL(L)}, \textit{Tacotron-PL(M)}, \textit{Tacotron-PL(H)}, and \textit{Tacotron-PL(LMH)}. As shown in Fig. \ref{fig:mos}, \textit{Tacotron-PL} models consistently outperforms \textit{Tacotron} baseline with either Griffin-Lim algorithm or WaveRNN vocoder, while \textit{Tacotron-PL(L)} achieves the best result.
Note that WaveRNN vocoder achieves better speech quality than Griffin-Lim algorithm, we conduct the subsequent listening experiments only with the speech samples generated by WaveRNN vocoder.




\begin{figure}[t]
    \centering
    \centerline{\includegraphics[width=80mm]{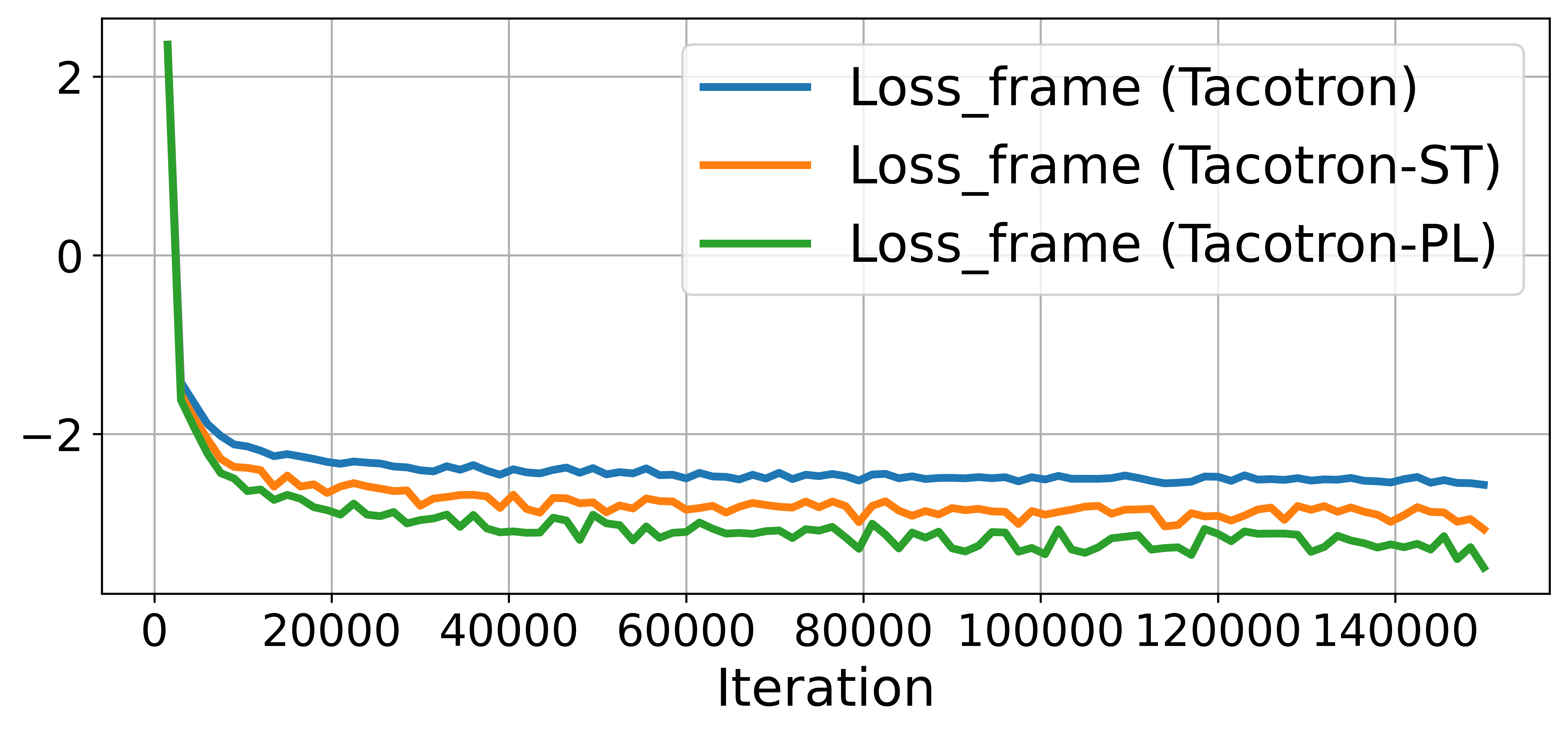}}
    \vspace{-2mm}
    \caption{The convergence trajectories of three loss values on LJ-Speech training data over the iteration steps, namely $Loss_{frame}$ for \textit{Tacotron} baseline, \textit{Tacotron-ST}, and  $Loss_{frame}$ component as part of the $Loss_{total}$ for \textit{Tacotron-PL}.
    }
    \vspace{-7mm}
    \label{fig:loss}
\end{figure}

\subsubsection{Expressiveness}

In the objective evaluations and MOS listening tests, \textit{Tacotron-PL(L)} and \textit{Tacotron-PL(LHM)} consistently offer better results. We next focus on comparing \textit{Tacotron-PL(L)} and \textit{Tacotron-PL(LHM)} with \textit{Tacotron} baseline. We first conduct the AB preference test to assess speech expressiveness of the systems. Each audio is listened by 15 subjects, each of which listens to 120 synthesized speech samples. Table \ref{tab:ab1} reports the speech expressiveness evaluation results. We note that \textit{Tacotron-PL(L)} outperforms both \textit{Tacotron} baseline and \textit{Tacotron-PL(LMH)} in the preference test. The results suggest that $\Psi_{low}$ is more effective than other deep style features to inform the speech style.

\subsubsection{Naturalness}
We further conduct the AB preference test to assess the naturalness of the systems. Each audio is listened by 15 subjects, each of which listens to 120 synthesized speech samples. Table \ref{tab:ab1} reports the naturalness evaluation results. Just like in the expressiveness evaluation, we note that \textit{Tacotron-PL(L)} outperforms both \textit{Tacotron} baseline and \textit{Tacotron-PL(LMH)} in the preference test. The results confirm that $\Psi_{low}$ is more effective to inform the speech style.

\subsubsection{Deep Style Features}

We finally conduct Best Worst Scaling (BWS) listening experiments to compare the four different \textit{Tacotron-PL} systems with different deep style features.
The subjects are invited to evaluate multiple samples derived from the different models, and choose the best and the worst sample. 
We perform this experiment for 18 different utterances, and each subject listens to 72 speech samples in total.
Each audio is listened by 15 subjects.  

Table \ref{tab:bws} summarizes the results. We can see that  \textit{Tacotron-PL(L)} is selected for 80\% of time as the best model and only 5\% of time as the worst model, that shows $\Psi_{low}$ is the most effective deep style features.

\vspace{-4mm}
\subsection{Comparison with GST-Tacotron paradigm}

We further compare \textit{Tacotron-PL} with the state-of-the-art expressive TTS framework, i.e., GST-Tacotron~\cite{wang2018style}.
The  original GST-Tacotron model \cite{wang2018style} is focused on style control and transfer, which differs from  \textit{Tacotron-PL}. For a fair comparison, we modify the GST-Tacotron framework and build a comparative system, denoted as \textit{Tacotron-ST}.
Specifically, the reference encoder of the GST-Tacotron model is replaced with a pre-trained SER-based style descriptor as described in Sec. \ref{subsec:pretraining}.
The style features $\Psi$ extracted by the reference encoder informs \textit{Tacotron-ST} the style information as GST-Tacotron does \cite{wang2018style}.
We then jointly train the whole \textit{Tacotron-ST} framework including the pre-trained SER-based reference encoder with $Loss_{frame}$.

\begin{figure}[t]
    \centering
    \centerline{\includegraphics[width=0.36\textwidth]{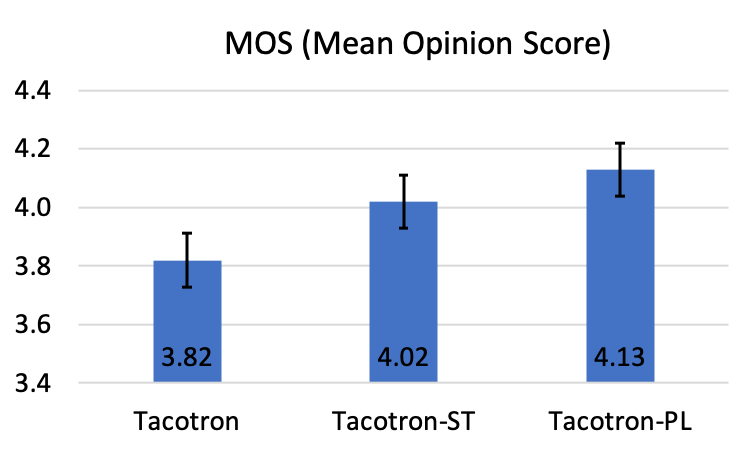}}
    \caption{The mean opinion scores (MOS) of three systems evaluated by 15 listeners, with 95\% confidence intervals computed from the t-test.}
    \label{fig:mos2}
\end{figure}

\begin{table}[]
\centering
\caption{The AB preference test for expressiveness and naturalness evaluation by 15 listeners, with 95\% confidence intervals computed from the t-test.}
\begingroup
\renewcommand{\arraystretch}{1} 
\begin{tabular}{p{4cm}p{0.7cm}p{0.7cm}p{0.7cm}<{\centering}p{0.8cm}<{\centering}}
\toprule
\multirow{2}{*}{\textbf{\tabincell{c}{Contrastive pair}
}} & \multicolumn{3}{l}{\textbf{Preference}(\%)} & \multirow{2}{*}{$\bm{p}$\textbf{-value}} \\ \cline{2-4}
 & \textbf{Former} & \textbf{Neutral} & \textbf{Latter} &  \\ \hline
 \multicolumn{5}{l}{\textbf{Expressiveness}}  \\ \hline
Tacotron vs. Tacotron-ST  & 30.22 & 20.44 & 49.34 &  0.00135\\
Tacotron-ST vs. Tacotron-PL  & 28.89 & 15.11 & 56.00 &  0.00129\\ \hline
\multicolumn{5}{l}{\textbf{Naturalness}} \\ \hline
Tacotron vs. Tacotron-ST  & 31.11 & 22.22 & 46.67 &  0.00133\\
Tacotron-ST vs. Tacotron-PL  & 30.67 & 20.89 & 48.44 &  0.00108\\ \bottomrule
\end{tabular}
\endgroup
\label{tab:ab2}
\end{table}

\textit{Tacotron-ST} and \textit{Tacotron-PL} share a similar architecture with Tacotron baseline~\cite{wang2018style} except that \textit{Tacotron-ST} is augmented by a reference encoder derived from a pre-trained SER model, while \textit{Tacotron-PL} is augmented by the proposed style reconstruction loss. 
In other words, both \textit{Tacotron-ST} and \textit{Tacotron-PL} incorporate style representations into the TTS training. 
We take \textit{Tacotron-ST} under the parallel style transfer scenario \cite{wang2018style} as the contrastive model for \textit{Tacotron-PL}. We also use the Tacotron model \cite{shen2018natural} as another baseline.

We use the low-level style feature $\Psi_{low}$ as the style embedding for \textit{Tacotron-ST} and the deep style feature for \textit{Tacotron-PL} in this section. We then conduct a set of experiments, following the previous experiment setup in Sec. \ref{subsec:expsetup}. 


\subsubsection{Convergence Trajectories of $Loss_{frame}$}

To examine the effect of the proposed training strategy, and the influence of and reference encoder and perceptual loss $Loss_{style}$,  we would like to observe how $Loss_{frame}$ converges with different training schemes on the same training data.  
We only compare the convergence trajectories of $Loss_{frame}$ between Tacotron baseline, \textit{Tacotron-ST} and the $Loss_{frame}$ component of $Loss_{total}$ for the training of \textit{Tacotron-PL} in Fig. \ref{fig:loss}.

A lower frame-level reconstruction loss, $Loss_{frame}$, indicates a better convergence, thus a better frame level spectral prediction. We observe that the $Loss_{frame}$ component in $Loss_{total}$ achieves a lower convergence value than $Loss_{frame}$ in traditional \textit{Tacotron} and {\textit{Tacotron-ST}} training. This suggests that utterance-level style objective function of \textit{Tacotron-PL} and reference signal supervision of \textit{Tacotron-ST} not only optimizes style reconstruction loss, but also reduces frame-level reconstruction loss over the \textit{Tacotron} baseline.

Finally, \textit{Tacotron-PL} obtains the best convergence trajectories during training, that further validates the proposed frame and style training strategy.
We note that the trajectories of \textit{Tacotron-PL(M)} vs. \textit{Tacotron-ST(M)}, \textit{Tacotron-PL(H)} vs. \textit{Tacotron-ST(H)}, \textit{Tacotron-PL(LMH)} vs. \textit{Tacotron-ST(LMH)} follow a similar pattern as \textit{Tacotron-PL(L)} vs. \textit{Tacotron-ST(L)}.

\subsubsection{Objective and Subjective Evaluation}

We also conduct objective and subjective evaluation experiments to compare the systems. 
In objective evaluation of \textit{Tacotron-ST}, we obtain 6.58, 1.14 and 14.18 of MCD, RMSE and FD respectively. The \textit{Tacotron-ST} results are consistently lower than those of \textit{Tacotron}, but higher than those of \textit{Tacotron-PL(L)}  in Table \ref{tab:mcd}, which further confirms the effectiveness of the frame and style training strategy.

In subjective evaluation, we conduct the MOS and AB preference tests to assess the overall performance of the systems. The MOS scores are reported in Fig. \ref{fig:mos2}.
Each audio is listened by 15 subjects, each of which listens to 75 synthesized speech samples. 
It is observed that \textit{Tacotron-PL} outperforms the \textit{Tacotron} and \textit{Tacotron-ST} baselines, that shows the clear advantage of frame and style training strategy. Table \ref{tab:ab2} reports the AB preference test results.
Each audio is listened by 15 subjects, each of which listens to 120 synthesized speech samples. 
All results show  that \textit{Tacotron-PL} outperforms both \textit{Tacotron} baseline and \textit{Tacotron-ST} significantly in terms of expressiveness and naturalness.

All the above experiments confirm that the proposed frame and style training strategy is more effective in informing the speech style than GST-Tacotron paradigm, which is encouraging.

\section{Conclusion}
We have studied a novel training strategy for Tacotron-based TTS system that includes frame and style reconstruction loss. We implement an SER model as the style descriptor to extract deep style features to evaluate the style reconstruction loss. We have conducted a series of experiments and demonstrated that the proposed Tacotron-PL training strategy outperforms the start-of-the-art Tacotron and GST-Tacotron-based baselines without the need of any add-on mechanism at run-time. While we conduct the experiments only on Tacotron, the proposed idea is applicable to  other end-to-end neural TTS systems, that will be the future work in our plan.

  

\begin{table}[th]
\centering
{APPENDIX A}
\caption {The scripts of utterances in six distinctive style groups from LJ-Speech database, the deep style features of which are visualized in Fig. 4.}
\label{tab:textgroup}
\resizebox{0.46\textwidth}{!}{%
\begin{tabular}{|c|l|}
\hline
\multirow{5}{*}{\begin{tabular}[c]{@{}c@{}} \textbf{Group 1} \\(Short Question)\end{tabular}} 
 
& (1) What did he say to that? \\
 & (2) Where would be the use? \\
 & (3) Where is it? \\
 & (4) The soldiers then? \\
 & (5) What is my proposal? \\
 
 \hline
\multirow{5}{*}{\begin{tabular}[c]{@{}c@{}} \textbf{Group 2} \\(Long Question)\end{tabular}} & \begin{tabular}[l]{@{}l@{}}(1) Could you advise me as to the general view we \\have on the American Civil Liberties Union?\end{tabular}\\
& \begin{tabular}[l]{@{}l@{}}(2) Why not relieve Newgate by drawing more largely \\upon the superior accommodation which Millbank offered?\end{tabular}\\
& \begin{tabular}[l]{@{}l@{}}(3) Who ever heard of a criminal being sentenced to \\catch the rheumatism or the typhus fever?\end{tabular}\\
& \begin{tabular}[l]{@{}l@{}}(4) Why not move the city prison bodily into this more\\ rural spot, with its purer air and greater breathing space?\end{tabular}\\
& \begin{tabular}[l]{@{}l@{}}(5) Great Britain in many ways has advanced further\\ along lines of social security than the United States?\end{tabular}\\
 \hline
\multirow{5}{*}{\begin{tabular}[c]{@{}c@{}} \textbf{Group 3} \\(Short Answer)\end{tabular}} & (1) Answer: Yes.\\
& (2) Answer: No.\\
& (3) Answer: Thank you.\\
& (4) Answer: No, sir.\\
& (5) Answer: By not talking to him.\\
 \hline
\multirow{5}{*}{\begin{tabular}[c]{@{}c@{}} \textbf{Group 4} \\(Short Statement)\end{tabular}} 
& (1) In September he began to review Spanish.\\
& (2) They agree that Hosty told Revill.\\
& (3) Hardly any one.\\
& (4) They are photographs of the same scene.\\
& (5) and other details in the picture.\\
 \hline
\multirow{5}{*}{\begin{tabular}[c]{@{}c@{}} \textbf{Group 5} \\(Long Statement)\end{tabular}} 
& \begin{tabular}[l]{@{}l@{}}(1) I only know that his basic desire was to get to Cuba \\by any means, and that all the rest of it was window \\dressing for that purpose. End quote.\end{tabular} \\
& \begin{tabular}[l]{@{}l@{}}(2) He tried to start a conversation with me several times,\\ but I would not answer. And he said that he didn't want \\me to be angry at him because this upsets him.\end{tabular} \\
& \begin{tabular}[l]{@{}l@{}}(3) Several of the publications furnished the Commission \\with the prints they had used, or described by\\ correspondence the retouching they had done.\end{tabular} \\
& \begin{tabular}[l]{@{}l@{}}(4) From an examination of one of the photographs, \\the Commission determined the dates of the issues  of \\the Militant and the Worker which Oswald was holding \\in his hand.\end{tabular} \\
& \begin{tabular}[l]{@{}l@{}}(5) He later wrote to another official of the Worker, \\seeking employment, and mentioning the praise he had\\ received for submitting his photographic work.\end{tabular} \\
 \hline
\multirow{5}{*}{\begin{tabular}[c]{@{}c@{}} \textbf{Group 6} \\(Digit String)\end{tabular}} & (1) Nineteen sixty-three.\\
& (2) Fourteen sixty-nine, fourteen seventy.\\
& (3) March nine, nineteen thirty-seven. Part one.\\
& (4) Section ten. March nine, nineteen thirty-seven. Part two.\\
& (5) On November eight, nineteen sixty-three.\\
 \hline
\end{tabular} 
}
\end{table}




\normalem
\bibliographystyle{IEEEtran}
{\footnotesize
\bibliography{refs}}

 
\begin{IEEEbiography}[{\includegraphics[width=1in,height=1.25in,clip,keepaspectratio]{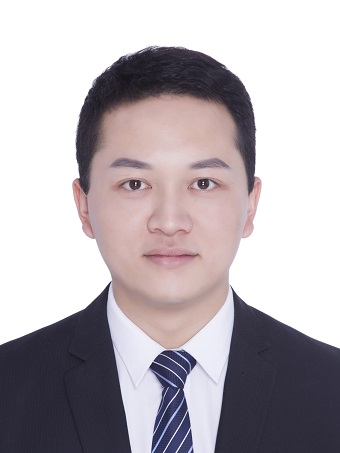}}]{Rui Liu}
received his B.S. degree from the Department of Software at Taiyuan university of technology, Taiyuan, China, in 2014. He is currently working toward
the Ph.D. degree in Inner Mongolia Key Laboratory of Mongolian Information Processing Technology, Inner Mongolia University, Hohhot, China. He is also an exchange PhD candidate at the Department of Electrical \& Computer Engineering of National University of Singapore, funded by China Scholarship Council (CSC). His research interests include prosody and acoustic modeling for speech synthesis, machine learning and natural language processing.
\end{IEEEbiography}

\begin{IEEEbiography}[{\includegraphics[width=1in,height=1.25in,clip,keepaspectratio]{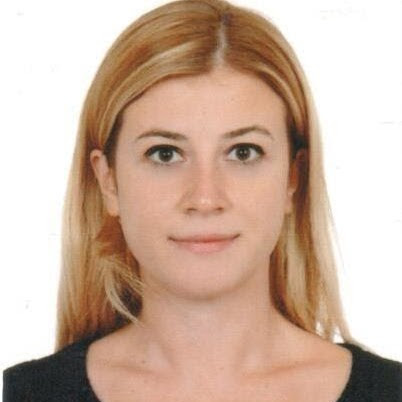}}]{Berrak Sisman}
received her PhD degree in Electrical and Computer Engineering from National University of Singapore in 2020, fully funded by A*STAR Graduate Academy under Singapore International Graduate Award (SINGA). She is currently working as an Assistant Professor at Singapore University of Technology and Design (SUTD). She is also an Affiliated Researcher at the National University of Singapore (NUS). Prior to joining SUTD, she was a Postdoctoral Research Fellow at the National University of Singapore, and a Visiting Researcher at Columbia University, New York, United States. She was also an exchange PhD student at the University of Edinburgh and a visiting scholar at The Centre for Speech Technology Research (CSTR), University of Edinburgh in 2019. She was attached to RIKEN Advanced Intelligence Project, Japan in 2018. Her research is focused on machine learning, signal processing, speech synthesis and voice conversion. She has served as the General Coordinator of the Student Advisory Committee (SAC) of International Speech Communication Association (ISCA). 
\end{IEEEbiography}

\begin{IEEEbiography}[{\includegraphics[width=1in,height=1.25in,clip,keepaspectratio]{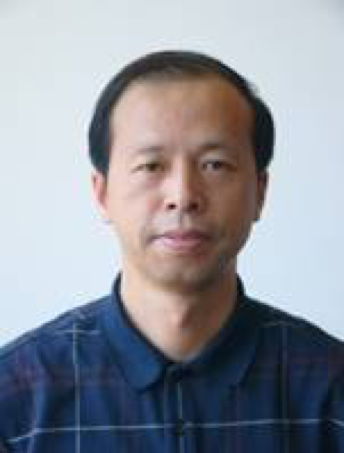}}]{Guanglai Gao}
received the B.S. degree from Inner Mongolia University, Hohhot, China, in 1985, and the
M.S. degree from the National University of Defense Technology, Changsha, China, in 1988. 
He was a Visiting Researcher at University of Montreal, Canada.
Currently, he is a Professor with the Department of Computer Science, Inner Mongolia University. His research interests include artificial intelligence and pattern recognition.
\end{IEEEbiography}
 
\begin{IEEEbiography}[{\includegraphics[width=1in,height=1.25in,clip,keepaspectratio]{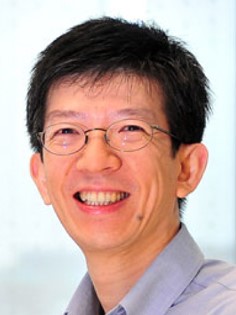}}]{Haizhou Li}
Haizhou Li (M’91-SM’01-F’14) received the B.Sc., M.Sc., and Ph.D degree in electrical and electronic engineering from South China University of Technology, Guangzhou, China in 1984, 1987, and 1990 respectively. Dr Li is currently a Professor at the Department of Electrical and Computer Engineering, National University of Singapore (NUS). His research interests include automatic speech recognition, speaker and language recognition, and natural language processing. Prior to joining NUS, he taught in the University of Hong Kong (1988-1990) and South China University of Technology (1990-1994). He was a Visiting Professor at CRIN in France (1994-1995), Research Manager at the Apple-ISS Research Centre (1996-1998), Research Director in Lernout \& Hauspie Asia Pacific (1999-2001), Vice President in InfoTalk Corp. Ltd. (2001-2003), and the Principal Scientist and Department Head of Human Language Technology in the Institute for Infocomm Research, Singapore (2003-2016). Dr Li served as the Editor-in-Chief of IEEE/ACM Transactions on Audio, Speech and Language Processing (2015-2018), a Member of the Editorial Board of Computer Speech and Language (2012-2018). He was an elected Member of IEEE Speech and Language Processing Technical Committee (2013-2015), the President of the International Speech Communication Association (2015-2017), the President of Asia Pacific Signal and Information Processing Association (2015-2016), and the President of Asian Federation of Natural Language Processing (2017-2018). He was the General Chair of ACL 2012, INTERSPEECH 2014 and ASRU 2019. Dr Li is a Fellow of the IEEE and the ISCA. He was a recipient of the National Infocomm Award 2002 and the President’s Technology Award 2013 in Singapore. He was named one of the two Nokia Visiting Professors in 2009 by the Nokia Foundation, and Bremen Excellence Chair Professor in 2019.

\end{IEEEbiography}
\enlargethispage{-9.5cm}
\end{document}